\documentstyle[aps,prl,twocolumn]{revtex}

\begin{document}
\draft
\title{ Exact solution of a spin-ladder model}
\author{Yupeng Wang}
\address{Institut f\"{u}r Physik, Universit\"at Augsburg, 86135 Augsburg,
Germany\\
and Laboratory of Ultra-Low Temperature Physics, Chinese Academy of Sciences, P. O. Box 2711, 
Beijing 100080, People's Republic of China}
\maketitle
\begin{abstract}
An integrable spin-ladder model with nearest-neighbor exchanges and biquadratic interactions is 
proposed. With the Bethe ansatz  solutions of the model hamiltonian, it is found that 
there are three possible phases in the ground
state, i.e., a rung-dimerized phase with a spin gap, and two massless phases. The 
possible fixed points of the system and the quantum critical behavior at the 
critical point $J=J_+^c$ are discussed.
\end{abstract}
Recently, there has been growing interest in the spin ladders for their relevance to 
 some quasi-one-dimensional materials, which under
hole-doping may show superconductivity\cite{1}. It is well known that the $S=1/2$
isotropic spin ladders with even number of legs have a spin-liquid ground state with an 
energy gap, while odd-legged ladders have a gapless spin-liquid ground state. On the other hand,
generalized ladders including other couplings beyond the nearest-neighbor exchanges, which 
can interpolate between a variety of systems, can show remarkably rich behavior.\cite{2,3,4}
In a recent paper\cite{2}, Nersesyan and Tsvelik predicted a new gapful phase for
the two-leg spin ladders, i.e., the dimerized phase driven by the four-spin interactions, which is
essentially different from the well known Haldane phase\cite{5}. This observation has been demonstrated
in a generalized spin-ladder model\cite{3} by constructing the exact ground state. Another
interesting phenomenon in the ladder systems is the quantum phase transition from the gaped
phase to the gapless phase, which has been studied experimentally in the Heisenberg ladder
$Cu_2(C_5H_{12}N_2)_2Cl_4$.\cite{6}\par
 As is well known, the 
integrable models provided us very good understanding to the correlated many-body systems in one 
dimension. However, a satisfactory integrable ladder model, which may play a similar role of the
Heisenberg chain\cite{7}, the one-dimensional (1D) Hubbard model\cite{8} and the supersymmetric
$t-J$ model\cite{9}, is still absent. The difficulty to construct an  integrable ladder model is 
almost the same we encountered in constructing a two-dimensional integrable model due to the strict
conditions for the integrability. For example, in 1D, there is only one path connecting two different
sites, while even for two coupled chains, we have a large number of paths connecting two different 
sites. We note that an integrable ladder model with artificial three-spin interactions has
been proposed recently\cite{10} and the integrability of a generalized spin ladder without
free parameter\cite{11} was addressed. The latter is more interesting but still defys Bethe-ansatz
solution. 
In this letter, we study a spin ladder with biquadratic interactions. By properly choosing
the four-spin coupling constants, we show the model is exactly solvable via algebraic Bethe ansatz. 
The model hamiltonian we shall study reads:
\begin{eqnarray}
H=\frac12J_1\sum_{j=1}^N[\vec{\sigma}_j\cdot\vec{\sigma}_{j+1}+\vec{\tau}_j\cdot\vec{\tau}_{j+1}]
+\frac12J_2\sum_{j=1}^N\vec{\sigma}_j\cdot\vec{\tau}_j\nonumber\\
+\frac14U_1\sum_{j=1}^N(\vec{\sigma}_j\cdot\vec{\sigma}_{j+1})
(\vec{\tau}_j\cdot\vec{\tau}_{j+1})\\+\frac14U_2\sum_{j=1}^N(\vec{\sigma}_j\cdot\vec{\tau}_j)
(\vec{\sigma}_{j+1}\cdot\vec{\tau}_{j+1}),\nonumber
\end{eqnarray}
where $\vec{\sigma}_j$ and $\vec{\tau}_j$ are Pauli matrices acting on site $j$ of the upper and lower legs,
 respectively;
$J_1$ and $J_2$ are the coupling constants along the legs and the rungs, respectively; $U_{1,2}$ are 
the 
biquadratic coupling constants and $N$ denotes the length of the ladder. Without the four-spin terms, 
Eq.(1)
represents the ordinary spin-ladder model. The new terms in Eq.(1) represent an interchain coupling and an
interrung coupling, which can be either effectively mediated by spin-phonon interaction or in the doped
phase generated by the conventional Coulomb repulsion between the holes moving in the spin correlated 
background as discussed in refs. [2,12]. The importance of biquadratic exchange for some properties of
$CuO_2$ plaquette has been pointed out\cite{13} and recent experiments revealed that such 
multi-spin-exchange interactions are realized in the two-dimensional (2D) solid $^3He$\cite{14}, 2D Wigner
solid of electrons formed in a $Si$ inversion layer\cite{15}, and the $bcc$ solid $^3He$\cite{16}.
 We note that when $U_2=0$, Eq.(1) is reduced to the model considered
in ref.[2]. For general parameters $J_{1,2}$ and $U_{1,2}$, 
the model 
(1) is still non-integrable. However, as we shall show below, when $U_1=J_1$, $U_2=0$ or $U_1=J_1$,
$U_2=-J_1/2$, the model is exactly solvable.
We shall study these cases through this paper. Not lossing generality, we  set $J_1=U_1=1$, $J_2
=J$ and $U_2=U$ in the following text. \par
We study first $U=0$ case. This is the simplest
integrable case but shows the main physics of the system. The hamiltonian (1) for $U=0$ can be 
rewritten as
\begin{eqnarray}
H=\frac14\sum_{j=1}^N(1+\vec{\sigma}_j\cdot\vec{\sigma}_{j+1})
(1+\vec{\tau}_j\cdot\vec{\tau}_{j+1})\nonumber\\
+\frac12J\sum_{j=1}^N(\vec{\sigma}_j\cdot\vec{\tau}_j-1)+\frac12(J-\frac12)N.
\end{eqnarray}
In this form, the integrability of the model is still somewhat hidden. To show it clearly, we note that
the first term in Eq.(2) can be rewritten as $\sum_{j=1}^N P_{j,j+1}$, where $P_{j,j+1}$ is the permutation
operator between two nearest rungs. Therefore, the first term of Eq.(2) is $SU(4)$-invariant as showed for
the spin-orbital model\cite{17}. An obvious fact is that $P_{j,j+1}$ can be expressed as $P_{j,j+1}=
\sum_{\alpha,\beta}X_j^{\alpha\beta}X_{j+1}^{\beta\alpha}$, where $X_{j}^{\alpha\beta}\equiv
|\alpha_j><\beta_j|$ are the Hubbard operators and the Dirac states $|\alpha_j>$ span the Hilbert 
space of the $j$-th rung
 and are orthogonal ($<\alpha_j|\beta_j>=\delta_{\alpha\beta}$). A basic representation of these
quantum states is $|\sigma_j^z,\tau_j^z>$, where $\sigma_j^z, \tau_j^z =\uparrow, \downarrow$. However, 
these states are not the eigenstates of the local operator $\vec{\sigma}_j\cdot\vec{\tau}_j$. This can
be overcome by choosing another basic representation
\begin{eqnarray}
|0>=\frac1{\sqrt 2}(|\uparrow,\downarrow>-|\downarrow,\uparrow>),\nonumber\\
|1>=|\uparrow,\uparrow>,\nonumber\\
|2>=\frac1{\sqrt 2}(|\uparrow,\downarrow>+|\downarrow,\uparrow>),\\
|3>=|\downarrow,\downarrow>.\nonumber
\end{eqnarray}
The first state denotes a singlet rung and the latter three indicate the spin-triplet states of a
rung. With these notations, Eq.(2) can be rewritten as (up to an irrelevant constant)
\begin{eqnarray}
H=\sum_{j=1}^N\sum_{\alpha,\beta=0}^3 X_j^{\alpha\beta}X_{j+1}^{\beta\alpha}-2J\sum_{j=1}^N X_j^{00}.
\end{eqnarray}
Obviously, the operators $N_\alpha\equiv\sum_{j=1}^NX_j^{\alpha\alpha}$,
which denote the numbers of the $\alpha$-rungs in the whole system, are conserved quantities. The 
constant $2J$ in the last term of Eq.(4) indicates a chemical potential applied on $N_0$ and reduces the
global $SU(4)$ symmetry of the hamiltonian to $U(1)\times SU(3)$. Now we have reduced the hamiltonian
(1) to an $SU(4)$-invariant spin chain (or equivalently an $SU(4)$ $t-J$ model), which can be 
solved by following the standard method\cite{9}. There are 
three branches of flavor waves (generalized spin waves) in this system. If we choose the reference 
state as
$|\Omega>=|0_1>\otimes|0_2>\otimes\cdots\otimes|0_N>$, these flavor waves describe the spin-triplet
``excitations". With this reference state, we obtain the Bethe-ansatz equations (BAE's):
\begin{eqnarray}
\left(\frac{\lambda_j-\frac i2}{\lambda_j+\frac i2}\right)^N=\prod_{l\neq j}^{M_1}\frac{\lambda_j-\lambda_l
-i}{\lambda_j-\lambda_l+i}\prod_{\alpha=1}^{M_2}\frac{\lambda_j-\mu_\alpha+\frac i2}{\lambda_j-\mu_\alpha
-\frac i2},\nonumber\\
\prod_{\beta\neq \alpha}^{M_2}\frac{\mu_\alpha-\mu_\beta-i}{\mu_\alpha-\mu_\beta+i}=\prod_{j=1}^{M_1}
\frac{\mu_\alpha-\lambda_j-\frac i2}{\mu_\alpha-\lambda_j+\frac i2}
\prod_{\delta=1}^{M_3}
\frac{\mu_\alpha-\nu_\delta-\frac i2}{\mu_\alpha-\nu_\delta+\frac i2},\\
\prod_{\gamma\neq\delta}^{M_3}\frac{\nu_\delta-\nu_\gamma-i}{\nu_\delta-\nu_\gamma+i}=\prod_{\alpha=1}^{M_2}
\frac{\nu_\delta-\mu_\alpha-\frac i2}{\nu_\delta-\mu_\alpha+\frac i2},\nonumber
\end{eqnarray}
where $M_1=N_1+N_2+N_3$, $M_2=N_2+N_3$ and $M_3=N_3$; $\lambda_j$, $\mu_\alpha$ and $\nu_\delta$ 
represent the rapidities of the flavor waves. Note the periodic boundary conditions $X_{N+1}^{\alpha\beta}
$ has been used in deriving Eq.(5). The eigenenergy (up to an irrelevant constant) 
of the hamiltonian (4) is given by 
\begin{eqnarray}
E=-\sum_{j=1}^{M_1}(\frac 1{\lambda_j^2+\frac14}-2J).
\end{eqnarray}
Obviously, for $0<J<2$, the ground state consists of real $\lambda_j$, $\mu_\alpha$ and 
$\nu_\delta$  closely packed around the origin. That means 
we have three ``Fermi seas" and three branches of gapless excitations. For $J>2$, the reference state
becomes the true ground state and any flavor excitation is gapful. The ground state is a product of 
the 
singlet rungs, which indicates the dimerization along the rungs. 
The energy gap can be easily deduced from Eq.(6) as $\Delta=2(J-2)$. $J_+^c=2$ 
indicates a quantum critical point at which the quantum phase transition from the dimerized phase to the
gapless phase occurs. At this critical point, all the three branches of flavor excitations are
marginal and the low-temperature thermodynamics of the system shows non-Fermi-liquid behavior as we
shall discuss below.  For $J<0$, the singlet rungs are unfavorable at low energy scales. For convenience,
we choose $|1_1>\otimes|1_2>\otimes\cdots\otimes|1_N>$ as the reference state. The BAE's are still given
by Eq.(5) but with $M_1=N_0+N_2+N_3$, $M_2=N_3+N_0$ and $M_3=N_0$. The eigenenergy (up to an irrelevant
constant) is given by
\begin{eqnarray}
E=-\sum_{j=1}^{M_1}\frac1{\lambda_j^2+\frac14}-2JN_0.
\end{eqnarray}
A hidden fact is that there is another critical value $J_-^c$. For $J<J_-^c$, no singlet rung exists
in the ground state configuration and the excitations consisting of singlet rungs are gapful. In 
this case, there are only two branches of gapless flavor waves and the effective low-energy hamiltonian
is equivalent to that of the $SU(3)$-invariant spin chain. The ground state consists of two ``Fermi
seas" (for $\lambda$ and $\mu$) with $M_1=2N/3$, 
$M_2=N/3$ and $M_3=0$. We denote the distributions of $\lambda$ and $\mu$ in the ground state as
$\rho_1(\lambda)$ and $\rho_2(\mu)$, respectively. A singlet excitation can be constructed by
introducing a $\nu$ mode and a $\mu$-hole $\mu_h$ in the BAE's. We denote
further the changes of $\rho_1(\lambda)$ and $\rho_2(\mu)$ via the $\nu$ mode and the $\mu$-hole
as $\delta\rho_1(\lambda)$
and $\delta\rho_2(\mu)$, respectively. From the BAE's (5) we can easily obtain
\begin{eqnarray}
\delta{\tilde\rho}_1(\omega)=\frac1{4\cosh^2\frac\omega 2-1}[e^{-i\nu\omega}-e^{-\frac12|\omega|}
e^{-i\mu_h\omega}],
\end{eqnarray}
where $\delta{\tilde\rho}_1(\omega)$ is the Fourier transformation of $\delta\rho_1(\lambda)$. Combining
Eq.(7) and Eq.(8), we derive the minimum energy (corresponding to $\nu\to0$, $\mu_h
\to\infty$)  to excite a $\nu$ mode from
the ground state as
\begin{eqnarray}
\epsilon_{min}=-\frac12\int \delta{\tilde\rho}_1(\omega)e^{-\frac12|\omega|}d\omega-2J\nonumber\\
=2|J|-\frac\pi{2
\sqrt3}+\ln\sqrt3.
\end{eqnarray}
The critical value $J_-^c$ is thus derived from $\epsilon_{min}=0$ as $J_-^c=-\pi/(4\sqrt3)+(ln3)/4$. 
For $J_-^c<J<0$, 
the system behaves as for $0<J<2$. Exactly at the critical point $J=J_-^c$, one branch 
of the flavor excitations (the singlet one) is marginal. Therefore, we have three quantum phases
in these system: A rung-dimerized phase when $J>J_+^c$, a gapless phase with three branches of
gapless flavor excitations when $J_+^c>J>J_-^c$ and another gapless phase with two branches of gapless
flavor excitations when $J<J_-^c$. We note that the dimerized phase shows a long range order
\begin{eqnarray}
<\Omega|X_i^{00}X_j^{00}|\Omega>=1,
\end{eqnarray}
which indicates the condensation of the singlet rungs. Under hole-doping, the system behaves as 
a $t-J$ ladder and the singlet rungs serve as Cooper pairs. The mobility of the singlet rungs under
hole-doping may drive the system to show superconductivity. Based on the above observations, we 
conclude that $J_\pm^c$ represent two unstable fixed points  of the system. In addition, the stable fixed
points of the system can be conjectured. For $J>J_+^c$, the transverse exchange dominates over the exchange
along the legs and the system should flow to a fixed point $J^*=+\infty$. For $J_-^c<J<J_+^c$, the two unstable
fixed points $J_\pm^c$ indicate an intermediate stable fixed point $J_-^c<J^*<J_+^c$.
For $J<J_-^c$, the singlet excitations are eliminated  at low
energy scales and the system should flow to a fixed point $J^*=-\infty$, which is equivalent to an
$SU(3)$-invariant spin chain. The gapless modes in the latter phase is mainly due to the high symmetry.
 Any small perturbations of $J_1$ or $U_1$ breaking this symmetry may drive it to the 
Haldane  phase as in the $SU(3)$-invariant spin-1 chain.
\par
Based on the BAE's, the thermodynamics of the present model can also be derived by following the standard
method\cite{18,19}. In the gapless phases, the system behaves as a Luttinger liquid\cite{20} and nothing
is anomalous. However, at the quantum critical points, the system may show non-Fermi-liquid behavior
due to the marginal excitations. We consider first the zero-temperature magnetic susceptibility for
$J=J_+^c$. Without the external field, the ground state is a condensate of singlet rungs. If we apply
a very weak external field on the system, some triplet rungs with $S_z=1$ appear in the ground-state
configuration while $N_2$ and $N_3$ still keep to be zero since the levels of these two types
of rungs are either lifted ($|3>$) or unchanged ($|2>$). 
 The energy density of the ground state in an external magnetic field ($H>0$)
reads
\begin{eqnarray}
E/N=\int_{-\Lambda}^\Lambda(4-\frac1{\lambda^2+\frac14}-H)\rho_1(\lambda)d\lambda,
\end{eqnarray}
where $\rho_1(\lambda)$ satisfies
\begin{eqnarray}
\rho_1(\lambda)+\int_{-\Lambda}^{\Lambda}a_2(\lambda-\lambda')\rho_1(\lambda')d\lambda'=a_1(\lambda),
\end{eqnarray}
with $a_n(\lambda)=n/2\pi[\lambda^2+(n/2)^2]$ and 
$\Lambda^2=1/(4-H)-1/4$.
For a small $H<<1$, we have $\Lambda\approx\sqrt{H}/4$ and Eq.(12) can be solved up to $O(H^{3/2},\lambda^2)$
as
\begin{eqnarray}
\rho_1(\lambda)=\frac2{\pi}-\frac1{\pi^2}\sqrt{H}+O(H^{3/2},\lambda^2).
\end{eqnarray}
Combining Eq.(13) and Eq.(11) we readily obtain the susceptibility as
\begin{eqnarray}
\chi=-\frac{\partial^2(E/N)}{\partial H^2}=\frac 1{2\pi}H^{-\frac12}+O(H^{\frac12}).
\end{eqnarray}
The low temperature susceptibility and the specific heat can also be derived from the so-called 
thermal Bethe ansatz.\cite{18,19} 
Via low-temperature expansion of the thermal BAE's\cite{21,22,23} we obtain
\begin{eqnarray}
C\sim T^{\frac12}, {~~~~~~~}\chi\sim T^{-\frac12},
\end{eqnarray}
which indicate a typical quantum critical behavior. These results can also be predicted by a 
simple flavor-wave theory with the dispersion relation $\epsilon(k)\sim k^2$. We note in the 
gaped phase, the magnetic field can also drive a quantum phase transition. At the quantum critical point
$H_c=2(J-2)$, similar quantum critical behavior can be obtained.
\par
Now we turn to the  $U=-1/2$ case. The last term in Eq.(1) can be rewritten with
the basic representation Eq.(3) as $-\sum_j[2X_j^{00}X_{j+1}^{00}-X_j^{00}+1/4]$.  
Up to an irrelevant constant, we rewrite Eq.(1) as
\begin{eqnarray}
H=\sum_{j=1}^N[\sum_{\alpha=1}^3(X_j^{\alpha0}X_{j+1}^{0\alpha}+X_j^{0\alpha}X_{j+1}^{\alpha0})
\nonumber\\
+\sum_{\alpha,\beta=1}^3X_j^{\alpha\beta}X_{j+1}^{\beta\alpha}-X_j^{00}X_{j+1}^{00}]
+(1-2J)\sum_{j=1}^N X_j^{00}.
\end{eqnarray}
The above hamiltonian deserves similarity to an $SU(1|3)$-supersymmetric t-J model, which still 
allows Bethe-ansatz solution. We choose still $|\Omega>$ as the reference state. 
The BAE's read:
\begin{eqnarray}
\left(\frac{\lambda_j-\frac i2}{\lambda_j+\frac i2}\right)^N=
\prod_{\alpha=1}^{M_2}\frac{\lambda_j-\mu_\alpha-\frac i2}{\lambda_j-\mu_\alpha
+\frac i2},\nonumber\\
\prod_{\beta\neq\alpha}^{M_2}\frac{\mu_\alpha-\mu_\beta-i}{\mu_\alpha-\mu_\beta+i}=\prod_{j=1}^{M_1}
\frac{\mu_\alpha-\lambda_j-\frac i2}{\mu_\alpha-\lambda_j+\frac i2}
\prod_{\delta=1}^{M_3}
\frac{\mu_\alpha-\nu_\delta-\frac i2}{\mu_\alpha-\nu_\delta+\frac i2},\\
\prod_{\gamma\neq\delta}^{M_3}\frac{\nu_\delta-\nu_\gamma-i}{\nu_\delta-\nu_\gamma+i}=\prod_{\alpha=1}^{M_2}
\frac{\nu_\delta-\mu_\alpha-\frac i2}{\nu_\delta-\mu_\alpha+\frac i2},\nonumber
\end{eqnarray}
The eigenenergy of the hamiltonian (16) is given by 
\begin{eqnarray}
E=\sum_{j=1}^{M_1}(\frac 1{\lambda_j^2+\frac14}+2J-1).
\end{eqnarray}
The situation is very similar to that of $U_1=1$, $U=0$ case. There are still three phases, i.e., 
a rung-dimerized phase and two gapless phases. For $J_-^c<J<1/2$, some triplet rungs are allowed in the ground
state. The ground-state configuration is described by closely packed real $\nu$ modes and the corresponding
$\lambda-3-$strings and $\mu-2-$strings:
\begin{eqnarray}
\lambda_\delta^n=\nu_\delta+i(2-n),{~~~~}n=1,2,3,\nonumber\\
\mu_\delta^\pm=\nu_\delta\pm\frac i2.
\end{eqnarray}
For $J>J_+^c=1/2$, we get again a dimerized ground state. Comparing to the $SU(4)$ case, we find $J_+^c$ is
remarkably reduced by a negative $U$. Notice that a negative $U$ indicates the attraction between two
nearest singlet-rungs. This attraction enhances the dimerization along the rung direction.
\par
For small positive $J$ or negative $J$, the triplet rungs are more stable than the singlet ones. 
For convenience, we choose $|1_1>
\otimes\cdots\otimes|1_N>$ as the reference state. The BAE's\cite{24} read
\begin{eqnarray}
\left(\frac{\lambda_j-\frac i2}{\lambda_j+\frac i2}\right)^N=\prod_{l\neq j}^{M_1}\frac{\lambda_j-\lambda_l
-i}{\lambda_j-\lambda_l+i}\prod_{\alpha=1}^{M_2}\frac{\lambda_j-\mu_\alpha+\frac i2}{\lambda_j-\mu_\alpha
-\frac i2},\nonumber\\
\prod_{\beta\neq\alpha}^{M_2}\frac{\mu_\alpha-\mu_\beta-i}{\mu_\alpha-\mu_\beta+i}=\prod_{j=1}^{M_1}
\frac{\mu_\alpha-\lambda_j-\frac i2}{\mu_\alpha-\lambda_j+\frac i2}
\prod_{\delta=1}^{M_3}
\frac{\mu_\alpha-\nu_\delta-\frac i2}{\mu_\alpha-\nu_\delta+\frac i2},\\
\prod_{\alpha=1}^{M_2}
\frac{\nu_\delta-\mu_\alpha-\frac i2}{\nu_\delta-\mu_\alpha+\frac i2}=1,\nonumber
\end{eqnarray}
where $M_1=N_2+N_3+N_0$, $M_2=N_3+N_0$ and $M_3=N_0$. The eigenenergy takes the same form of Eq.(7)
but with $J\to J-1/2$. $J_-^c$ can be easily derived from Eq.(20) as
$J_-^c=\frac12-\frac\pi{4\sqrt3}+\frac14\ln3$.
Interestingly, $J_-^c$ takes a positive value in this case. The space of the phase with three branches 
of massless excitations is remarkably
depressed by the attractive rung-rung interaction. This observation strongly indicates that there is
a critical point $U=U_c$. When $U<U_c$, $J_\pm^c$ coincide each other and the intermediate fixed point
will be eliminated, implying only two phases can exist in the system.\par
In conclusion, we propose an integrable spin-ladder model which exhibits rich physics. This
model may play a similar role in the spin-ladder systems as the supersymmetric $t-J$ model does in the
one-dimensional correlated electron systems.\par
The author acknowledges valuable communications with S.M. Fei. He is also indebted to the hospitality
of Institut f\"{u}r Physik, Universit\"{a}t Augsburg. This work was partially supported by
AvH-Stiftung, the key projects of National Natural Science Foundation of China, the key projects
of Chinese Academy of Sciences and the OYS-Foundation of China.

\end{document}